\documentclass[aps,prd,a4paper,twocolumn,amsmath,showpacs,superscriptaddress,nofootinbib,preprintnumbers]{revtex4-2}
\usepackage{graphicx}
\usepackage{epsf}
\usepackage{bm}
\usepackage{amsmath}
\usepackage{amsfonts}
\usepackage{amssymb}
\usepackage{epstopdf}
\usepackage{natbib}
\usepackage[dvipsnames,table]{xcolor}
\usepackage{physics}
\usepackage{hyperref}
\usepackage{lipsum}
\usepackage{float}
\usepackage{multirow}

\usepackage[capitalize]{cleveref}
\providecommand{\U}[1]{\protect\rule{.1in}{.1in}}
\usepackage[normalem]{ulem}
\usepackage[dvipsnames]{xcolor}
\hypersetup{colorlinks,linkcolor={magenta},citecolor={red},urlcolor={blue}}

\newcommand{\be}{\begin{equation}}
\newcommand{\ee}{\end{equation}}
\newcommand{\bea}{\begin{eqnarray}}
\newcommand{\eea}{\end{eqnarray}}

\begin{document}


\title{Testing an oscillatory behavior of dark energy}

\author{Luis A. Escamilla}
\email{luis.escamilla@icf.unam.mx}
\affiliation{School of Mathematics and Statistics, University of Sheffield, Hounsfield Road, Sheffield S3 7RH, United Kingdom}
\affiliation{Instituto de Ciencias F\'isicas, Universidad Nacional Aut\'onoma de M\'exico, Cuernavaca, Morelos, 62210, M\'exico}

\author{Supriya Pan}
\email{supriya.maths@presiuniv.ac.in}
\affiliation{Department of Mathematics, Presidency University, 86/1 College Street,  Kolkata 700073, India}
\affiliation{Institute of Systems Science, Durban University of Technology, PO Box 1334, Durban 4000, Republic of South Africa}

\author{Eleonora Di Valentino}
\email{e.divalentino@sheffield.ac.uk}
\affiliation{School of Mathematics and Statistics, University of Sheffield, Hounsfield Road, Sheffield S3 7RH, United Kingdom}

\author{Andronikos Paliathanasis}
\email{anpaliat@phys.uoa.gr}
\affiliation{Institute of Systems Science, Durban University of Technology, PO Box 1334, Durban 4000, Republic of South Africa}
\affiliation{Departamento de Matem\'{a}ticas, Universidad Cat\'{o}lica del Norte, Avda.
Angamos 0610, Casilla 1280 Antofagasta, Chile}

\author{Jos\'e Alberto V\'azquez}
\email{javazquez@icf.unam.mx}
\affiliation{Instituto de Ciencias F\'isicas, Universidad Nacional Aut\'onoma de M\'exico, Cuernavaca, Morelos, 62210, M\'exico}

\author{Weiqiang Yang}
\email{d11102004@163.com}
\affiliation{Department of Physics, Liaoning Normal University, Dalian, 116029, P. R. China}

\begin{abstract}
The main aim of this work is to use a model-independent approach, along with late-time observational probes, to reconstruct the Dark Energy (DE) equation of state $w_{\rm DE}(z)$. Our analysis showed that, for a late time universe, $w_{\rm DE}$ deviates from being a constant but in contrast exhibits an oscillatory behavior, hence both quintessence ($w_{\rm DE}> -1$) and phantom ($w_{\rm DE} < -1$) regimes are equally allowed. In order to portray this oscillatory behavior, we explored various parametrizations for the equation of state and identified the closest approximation based on the goodness of fit with the data and the Bayesian evidence analysis. Our findings indicated that while all considered oscillating DE parametrizations provide a better fit to the data, when compared to the cosmological constant, they are penalized in the Bayesian evidence analysis due to the additional free parameters. Overall, the present article demonstrates that in the low redshift regime, the equation of state of the DE prefers to be dynamical and oscillating.  We anticipate that future cosmological probes will take a stand in this direction. 
\end{abstract}
\maketitle

\section{Introduction}

The physics of the dark sector in our universe, consisting of two dominant components (which together make up nearly 96\% of the total energy budget), namely, a non-luminous Dark Matter (DM) fluid and Dark Energy (DE), has remained a significant mystery for cosmologists for the past two decades. The DM sector plays a crucial role in the observed structure formation of our universe, while the DE is believed to be responsible for the unknown mechanism driving the current accelerating phase of our universe. Based on existing observational evidence, DM behaves as a pressure-less or cold fluid (referred to as CDM), while DE is adequately described by a positive cosmological constant, $\Lambda$, incorporated into the gravitational equations of Einstein's General Relativity. 
The cosmological constant corresponds to the vacuum energy with a constant equation of state (EoS)  $w = p_{\Lambda}/\rho_{\Lambda} = -1$, where $p_{\Lambda}$ and $\rho_{\Lambda}$ are the pressure and energy density of the vacuum energy sector, respectively. In this case the DE is non-dynamical.  The resulting picture $-$  known as $\Lambda$CDM or standard model cosmology $-$ has been quite successful in explaining  a large span of the observational datasets.

However, $\Lambda$CDM cosmology is facing both theoretical and observational challenges. 
The cosmological constant problem~\cite{Weinberg:1988cp} and the cosmic coincidence problem~\cite{Zlatev:1998tr} were among the initial challenges for $\Lambda$CDM, and motivated cosmologists to explore alternative scenarios. On the other hand, recent advancements in observational data consistently report discrepancies in the estimations of key cosmological parameters obtained from early-time measurements, such as those from the cosmic microwave background radiation analyzed by Planck (under the assumption of the $\Lambda$CDM paradigm), when compared to estimations from other astronomical probes. Specifically, the $\sim 5\sigma$ tension on the Hubble constant $H_0$ between the $\Lambda$CDM based Planck~\cite{Planck:2018vyg} 
and the  SH0ES (Supernova $H0$ for the Equation of State) collaboration~\cite{Riess:2021jrx} has emerged as a significant issue and probably demands a revision of the standard cosmological model.

Given the impressive fit of $\Lambda$CDM with numerous astronomical probes, one may expect that this model might be an approximate version of the ultimate cosmological picture that is yet to be unveiled.  As a consequence, several revisions of it have been suggested and they have been consequently investigated using various astronomical probes~\cite{Verde:2019ivm,Knox:2019rjx,DiValentino:2020zio,Jedamzik:2020zmd,DiValentino:2021izs,Abdalla:2022yfr,Kamionkowski:2022pkx,Khalife:2023qbu}. In particular in the DE sector, among the others in the late time universe, revisions can be categorized into two main directions: DE scenarios in the context of Einstein's General Relativity~\cite{Copeland:2006wr,Bamba:2012cp,Tutusaus:2023cms} and alternative models based on modified gravity~\cite{Nojiri:2006ri,Sotiriou:2008rp,DeFelice:2010aj,Clifton:2011jh,Cai:2015emx,Nojiri:2017ncd,Bahamonde:2021gfp}.
Focusing on the former, the simplest revision of a $\Lambda$CDM cosmology appears in the introduction of a dynamical DE, where the DE EoS, 
$w_{\rm DE} = p_{\rm DE}/\rho_{\rm DE}$, is either a constant with a value different from $-1$, or, it is a function of the cosmic variables.   
Since there is no fundamental principle to derive $w_{\rm DE}$, therefore, a variety of parametrized forms of $w_{\rm DE}$ have been  proposed in the literature~ \cite{Efstathiou:1999tm,Chevallier:2000qy,Astier:2000as,Weller:2001gf,Linder:2002et,Wetterich:2004pv,SDSS:2004kqt,Jassal:2004ej,Alam:2004jy,Upadhye:2004hh,Gong:2005de,Linder:2005ne,Nesseris:2005ur,Barboza:2008rh,Barboza:2009ks,Ma:2011nc,Sendra:2011pt,Li:2011dr,DeFelice:2012vd,Feng:2012gf,Novosyadlyj:2013nya,Magana:2017usz,Rezaei:2017yyj,Wang:2017lai,Yang:2017alx,Yang:2018prh,Mehrabi:2018oke,Yang:2018qmz,Li:2019yem,Pan:2019brc,Hernandez-Almada:2020uyr,Benaoum:2020qsi,Li:2020ybr,Yang:2021flj,Yang:2021eud,Escamilla:2023oce,Rezaei:2023xkj,Reyhani:2024cnr}.  This is one of the potential approaches in which observational data are used to restrict the choices of the dynamical DE models. On the contrary, one can reconstruct the evolution of $w_{\rm DE}$ without making any prior assumptions about the underlying model. This alternative approach is appealing because it leads to a model-independent reconstruction of the DE EoS, and, because of this characteristic, it could be used to discriminate between the dark energy parametrizations. 
Recognizing the potential nature of this approach in discriminating between the DE parametrizations based on their EoS, we conducted a model-independent 
reconstruction of $w_{\rm DE}$ using some well known and most used astronomical probes. Our analysis revealed that the reconstructed $w_{\rm DE}$ exhibits an oscillatory behavior (in the following sections, we will elaborate on the methodology of this reconstruction and present corresponding plots to illustrate our findings) with the evolution of the universe.

The possibility of an oscillating behavior of $w_{\rm DE}$ was also observed in some earlier works~\cite{Zhao:2017cud,Zhang:2019jsu} where the authors reconstructed the DE EoS following a non-parametric approach and using various observational datasets. 
Before the observational hints for an oscillating DE EoS, existence of periodic behavior in some cosmological models was noticed.   
In the context of single scalar field cosmologies, it was found that a scalar field potential, which is minimally coupled to gravity, can induce an oscillating behaviour in the expansion rate of the universe~\cite{Rubano:2003er}, and as a consequence, this oscillatory behavior can be transmitted (through the Friedmann equations) to the energy density of the scalar field and its EoS. On the other hand, multi-scalar field models~\cite{Lazkoz:2007mx,Brown:2017osf,Paliathanasis:2020wjl} and modified theories of gravity (see for instance~\cite{Xu:2012jf,Leon:2012mt,Paliathanasis:2022xoq}) can also instigate oscillatory nature in the associated dynamical variables. Specifically, within these frameworks, the periodic behaviour of the cosmological parameters exists, corresponding to asymptotic solutions described by the saddle critical points of the associated autonomous dynamical systems. It was argued in Ref.~\cite{Linder:2005dw} that oscillating DE can solve the cosmic coincidence problem. 
These altogether motivated several investigators to work with the oscillating DE EoS~\cite{Feng:2004ff,Nojiri:2006ww,Kurek:2007bu,Jain:2007fa,Saez-Gomez:2008mkj,Kurek:2008qt,Pace:2011kb,Pan:2017zoh,Panotopoulos:2018sso,Tamayo:2019gqj,Colgain:2021pmf,Rezaei:2019roe,Guo:2022dno,Rezaei:2024vtg}.


As the evolution of DE is not yet uncovered, therefore, in principle, the oscillation in the DE EoS cannot be excluded unless the observational data is in strong disagreement with such possibility.  And the reconstructed $w_{\rm DE}$ as done in the present article together with similar findings in the earlier works \cite{Zhao:2017cud,Zhang:2019jsu}, further strengthen the possibility of oscillations in the DE EoS. Our model-independent reconstruction, along with this theoretical background, will serve as motivation for us to propose a number of new DE parameterizations, designed in such a way that they present oscillatory behavior at late times with "as minimal as possible" a number of free parameters. 

The article is organized as follows. In section~\ref{sec-osc} we describe the gravitational equations and introduce the oscillating DE parametrizations.  In section~\ref{sec-data},
we describe the observational data and methodology. Then in section~\ref{sec-results} we discuss the results and their implications. Finally, in section~\ref{sec-summary}, we give our conclusions.

\section{Oscillating dark energy}
\label{sec-osc}

We consider that the geometry of our universe is well described by the Friedmann-Lema\^{i}tre-Robertson-Walker (FLRW) line element given by 

\begin{equation}
ds^2 = -dt^2 + a^2(t) \left[\frac{dr^2}{1- Kr^2} + r^2 (d\theta^2 + \sin^2 \theta d\phi^2)\right],
\end{equation}
in which $(t, r, \theta, \phi)$ are the co-moving coordinates; $a (t)$ is the expansion scale factor of the universe (hereafter denoted simply by ``$a$") which is related to the redshift $z$ as $1+z = a_0/a (t)$ ($a_0$ is the present value of the scale factor); $K$ stands for the spatial geometry of the universe where $K =0$ corresponds to a spatially flat universe; $K =-1$ corresponds to an open universe, and $K = 1$ corresponds to a closed universe.    
We assume that the gravitational sector of the universe is described by Einstein's General Relativity and the matter sector of the universe is minimally coupled to gravity. In this framework, one can write down the Einstein's gravitational equations as follows 
\begin{eqnarray}
&& \rho = \frac{3}{8 \pi G} \left( H^2 + \frac{K}{a^2} \right),\label{fe-1}\\
&& \frac{\ddot{a}}{a} = - \frac{4 \pi G}{3} (3 p + \rho), \label{fe-2}
\end{eqnarray}
where an overhead dot represents the derivative with respect to the cosmic time; $H \equiv \dot{a}/a$ is the Hubble rate of the FLRW universe; $p$ and $\rho$ are, respectively, the pressure and energy density of the total matter sector of the universe. 
From equations (\ref{fe-1}) and (\ref{fe-2}), or alternatively, using the Bianchi's identity, one arrives at the conservation equation of the total fluid 
\begin{eqnarray}
  \dot{\rho} + 3 H (p +\rho) = 0.   
\end{eqnarray}
As $\rho$ includes the total energy density of all the fluids and if the fluids do not interact with each other (which is the case of this article), hence, for the $i$-th fluid, the conservation equation becomes $\dot{\rho}_i + 3 H (p_i +\rho_i) = 0$.  
Now, as usual, we consider that the matter sector of the universe is comprised of pressure-less matter (baryons+cold DM) and a DE fluid. We have neglected radiation as its effects are negligible in the late time. 
All the fluids obey a barotropic EoS $p_i = w_i \rho_i$ where for pressure-less matter $w_{\rm m} = 0$ and for DE, $w_{\rm DE} (a)$ is dynamical. 
From the conservation equation of each fluid, one can derive that $\rho_{\rm r}  = \rho_{{\rm r}0} (a/a_0)^{-4}$, $\rho_{\rm m} = \rho_{{\rm m}0} (a/a_0)^{-3}$, where $\rho_{{\rm r}0}$ and $\rho_{{\rm m}0}$ are respectively the present values of $\rho_{\rm r}$ and $\rho_{\rm m}$. 
Finally, the evolution of the DE density becomes:
\begin{eqnarray}
\rho_{\rm DE} = \rho_{\rm DE, 0}\; \exp(-\int_{a_0}^{a} \frac{1+w_{\rm DE} (a)}{a}~da), 
\end{eqnarray}
which dictates that depending on the functional form of $w_{\rm DE}$, the evolution of the energy density varies. 
In this work, we assume that $w_{\rm DE} (a)$ has an oscillating nature. As one can propose a variety of oscillating DE EoS parametrizations, in this work we started with some already known ones with many free parameters and then considered some other new parametrizations with fewer free parameters.   
In Table~\ref{tab:models}, we summarize the EoS parametrizations that we wish to study. Some of them have already been proposed in the literature, and some of them are new, proposed in this work. The new parameterizations in this work (models 8-12) are proposed as a way to reproduce a heavy oscillatory behavior at late times ($0<z<3$) with as few parameters as possible (2 in this case).
Our five proposed parameterizations also share the important feature that, at large redshifts (early times), they recover the behavior $w_{\rm DE}(z) \approx -1$. This is clearly seen by the terms $\frac{1}{1+z^2}$ and $\frac{z}{1+z^2}$, which are factors in the oscillatory part of the equation and tend to zero as $z$ becomes sufficiently large. This behavior is desirable for achieving a $\Lambda$CDM-like evolution in the early universe, while allowing for an oscillating EoS for dark energy (DE) at later times. In models 1, 3, 4, 5, and 7, where the oscillatory component does not have a mechanism for diminishing amplitude as $z$ increases, this early-time behavior is absent. On the other hand, models 2 and 6 exhibit $w(z) \approx w_1$ and $w(z) \approx w_0$ as $z \rightarrow \infty$, respectively. To replicate the standard model's early-time behavior in these cases, it would be necessary for $w_1$ and $w_0$ to approach $-1$.
We would like to clarify that this does not mean that the proposed parameterizations are the only available ones to exhibit these features, but rather that we choose to focus on them. Other options are possible, but their study is beyond the scope of this work.

Let us note that,  given a particular choice of the oscillating  $w_{\rm DE}$, one can further investigate how this affects other cosmic variables. For example, the deceleration parameter $q = - (1+ \dot{H}/H^2)$ can be expressed in terms of the cosmic fluids as follows:

 \begin{eqnarray}\label{deceleration}
     q = -1 - \frac{K}{a^2 H^2} + 4 \pi G \left(\frac{p+\rho}{H^2} \right).
 \end{eqnarray}
Now, for a specific EoS of DE, one can trace the evolution of the universe
from eqn. (\ref{deceleration}) and investigate whether for this particular EoS of DE, our universe enters into the accelerating phase from the decelerating one. In this article we have considered the spatial flatness of the FLRW universe, i.e., $K =0$ and without any loss of generality we set $a_0 =1$.

\begin{table*}[]
    \centering
    \caption{The table summarizes a variety of oscillating dark energy parametrizations considered in this work. It is very important to note that in all the parametrizations displayed above, $w_0$ does not always refer to the present value of the DE EoS. We have kept similar notations as used by the corresponding authors. Thus, the present value of the DE EoS is the one that can be obtained by setting $z = 0$ in $w_{\rm DE} (z)$. For example, in Models 4 $-$ 7, $w_0$ is the present value of the DE EoS, but in Model 1, 2, 3, the present value of the DE EoS are, $ w_0 + w_1  \cos (A \ln (1/a_c))$, $w_1 + w_2\; \cos \left(w_3 + a_c\right)$ and $w_c - A \sin (\theta)$, respectively. On the other hand, for Models 8 $-$ 12, the present value of the DE EoS is $-1$.  }
    \begin{tabular}{ccccccccc}
    \hline 
        Model No.  &   Expression of $w_{\rm DE}$ & Free parameters & Reference\\ 
        \hline

 1  & $ w_0 + w_1  \cos (A \ln (a/a_c)) $ & 4 & Ref.~\cite{Feng:2004ff}  \\\\

 2 &  $w_1 + w_2 a\; \cos \left(w_3 a + a_c\right)$  & 4  &  Ref.~\cite{Zhao:2005vj}  \\\\

3 &   $w_c - A \sin (B \ln a + \theta)$ & 4 &  Ref.~\cite{Linder:2005dw} \\\\
 
4   & $w_0 + b \left\{  1- \cos \left[\ln (1+z)  \right] \right\}$  & 2   &   Ref.~\cite{Pan:2017zoh}  \\\\

5  &   $w_0 + b \sin \left[   \ln (1+z)  \right]$ & 2 &   Ref.~\cite{Pan:2017zoh} \\\\

6 &  $w_0 + b \left[  \frac{\sin (1+z)}{1+z} - \sin 1\right]$  & 2  &   Ref.~\cite{Pan:2017zoh} \\\\

7 &   $w_0 + b \left[ \frac{z}{1+z} \right] \cos(1+z)$ & 2 & Ref.~\cite{Pan:2017zoh} \\\\

8  &  $-1 + \frac{w_1}{1+z^2} \sin (w_2\, z)$ &  2 &  This work \\\\

9  &  $-1 + \frac{w_1 z}{1+z^2} \sin \left(w_2 z\right)$ & 2 &  This work \\\\

10 & $-1 + \frac{w_1 z}{1+z^2} \cos \left(w_2 z\right)$ & 2 &  This work \\\\

11 &  $-1 + \frac{w_1 z}{1+z^2} (\cos \left(w_2 z\right))^2$ & 2 & This work \\\\

12 &  $-1 + \frac{w_1 z}{1+z^2} (\cos \left(w_2 z\right))^3$ & 2 & This work \\

\hline 
    \end{tabular}
    \label{tab:models}
\end{table*}

\section{Observational data and the methodology}
\label{sec-data} 

As our objective is to carry out parameter inference within various DE models, it is imperative to have suitable datasets at our disposal. In this study, we will utilize the following ones:

\begin{itemize}
    \item The \textit{PantheonPlus} supernova type Ia (SNeIa) sample~\cite{Scolnic:2021amr}, which encompasses 1701 light curves corresponding to 1550 unique SNeIa  within the redshift range of $0.001<z<2.26$. This dataset will be referred to as \textbf{SN}.
  
    \item Measurements of Baryon Acoustic Oscillation (BAO) containing the SDSS Galaxy Consensus, quasars and Lyman-$\alpha$ forests~\cite{eBOSS:2020yzd}. The sound horizon is calibrated by using BBN~\cite{Cooke:2013cba}. These datasets are thoroughly detailed in Table 3 of Ref.~\cite{eBOSS:2020yzd}. In this work, we will collectively refer to this set of measurements as \textbf{\textit{BAO}}.
    
    \item The Hubble parameter serves as a key indicator of the Universe's expansion rate. This parameter is determined by collecting measurements from ancient stars, referred to as cosmic chronometers, which effectively act as ``standard clocks'' in cosmology. In our study, we rely on a compilation of 31 of these cosmic chronometers~\cite{jimenez2003constraints, simon2005constraints, stern2010cosmic, moresco2012new, zhang2014four, moresco2015raising, moresco20166}, denoted as \textbf{\textit{CC}} in the datasets. This dataset and its covariance matrix are available within the sampler code used in this work\footnote{ One can access them from the publicly available \texttt{SimpleMC}'s repository \url{https://github.com/ja-vazquez/SimpleMC/tree/master/simplemc/data}.}
    and this dataset is explicitly shown in Table 1 of~\cite{Yu:2017iju}.


\end{itemize}

\begin{figure}
    \centering
    \includegraphics[width=0.55\textwidth]{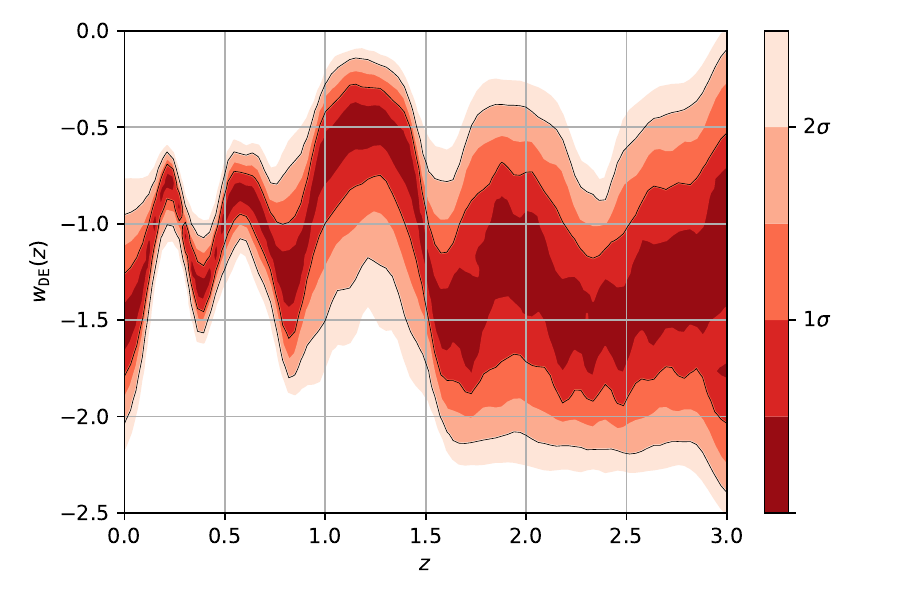}
    \caption{The reconstructed EoS for DE for 20 bins using the combined dataset from  CC+SN+BAO reveals an oscillating feature of $w_{\rm DE}$, particularly at low redshifts.  }
    \label{fig:d2comp}
\end{figure}

To determine the optimal parameter values for our models, we employ a customized Bayesian inference code, known as SimpleMC~\cite{simplemc, aubourg2015cosmological}, specifically designed for computing expansion rates and distances based on the Friedmann equation. For each model, we calculate its Bayesian evidence denoted as $E_i$. To facilitate a comparison between two different models (designated as 1 and 2), we rely on the Bayes' factor $B_{1,2}$, defined as the ratio of their respective evidences, or more precisely, the natural logarithm of this ratio. When using the empirical revised Jeffreys' scale, shown in \cref{jeffreys}, we gain a robust understanding of the relative performance of alternative models.

\begin{table}[t!]
\caption{Revised Jeffreys' scale for model selection. }
\footnotesize
\scalebox{1}{%
\begin{tabular}{cccc} 
\cline{1-4}\noalign{\smallskip}
 \vspace{0.15cm}
 $\ln{B_{12}}$ & Odds  &   Probability  &  Strength of evidence \\
\hline
\hline
\vspace{0.15cm}
$<$ 1.0 & $<$3:1 & $<$0.75 & Inconclusive  \\
\vspace{0.15cm}
1.0 &  $\sim$3:1 & 0.750 &   Weak evidence  \\
\vspace{0.15cm}
2.5 & $\sim$12:1  &  0.923  & Moderate evidence  \\
\vspace{0.15cm}
5.0 & $\sim$150:1 &  0.993  & Strong evidence  \\
\hline
\hline
\end{tabular}}
\label{jeffreys}
\end{table}

In assessing the goodness of fit of our reconstructions, particularly in relation to $\Lambda$CDM, we employ the value $-2\ln \mathcal{L_{\rm max}}$ for each model, where $\mathcal{L_{\rm max}}$ is the maximum likelihood obtained within a Bayesian context. For a comprehensive review of cosmological Bayesian inference, we refer to~\cite{Padilla:2019mgi}. The SimpleMC code also makes use of the \texttt{dynesty} library~\cite{speagle2020dynesty}, which leverages nested sampling techniques to compute the Bayesian evidence. The selection of the number of live points adheres to the general rule of thumb, specifically $50 \times \mbox{ndim}$~\cite{dynestyy}, where $\mbox{ndim}$ denotes the total number of parameters subject to sampling.

As stated in the introduction, the first reconstruction made consists of a model independent approach using a binning scheme. In this method, steps or bins are used to represent any function $f$, connecting them with hyperbolic tangents to ensure continuity. The target function is defined as follows:
\begin{equation}
f(z)= f_1+\sum^{N-1}_{i=1}\frac{f_{i+1} - f_i}{2}\bigg(1+\tanh{\Big(\frac{z-z_i}{\xi}\Big)} \bigg),
\label{bin_equation}
\end{equation}
where $N$ is the number of bins, $f_i$ is the amplitude of the bin value, $z_i$ is the position where the bin begins in the $z$ axis, and $\xi$ is a smoothness parameter. For a detailed explanation on how this approach to a reconstruction works please refer to~\cite{Escamilla:2021uoj}. 
In this case the reconstructed quantity is the EoS parameter $w_{\rm DE}(z)$, so the bins take the form of $w_i$. The number of bins is $N=20$ and the smoothness parameter takes the value of $\xi=0.1$. The flat priors will be $w_i(z_i): [-2.5, 0.0]$.

Regarding the flat priors of the cosmological parameters for every reconstruction, including the EoS parametrizations, the following ranges were imposed: $\Omega_m$ (matter density parameter) was constrained within the range of $[0.1, 0.5]$, $\Omega_b h^2$ (physical baryon density) within $[0.02, 0.025]$, and $h$ (dimensionless Hubble parameter, where $H_0 = 100h$ s$^{-1}$ Mpc$^{-1}$ km) within $[0.4, 0.9]$. For the model-specific parameters of the oscillating DE parametrizations we choose the following flat priors: 
\begin{itemize}
    \item \textbf{Model 1}: $w_0=[-2.0, 0.0]$, $w_1=[-5.0, 5.0]$, $A=[-5.0,5.0]$, $a_c=[0.001,10]$
    \item \textbf{Model 2}: $w_1=[-2.0, 0.0]$, $w_2=[-5.0, 5.0]$, $w_3=[-5.0,5.0]$, $a_c=[0.001,10]$
    \item \textbf{Model 3}: $w_c=[-2.0, 0.0]$, $A=[-5.0, 5.0]$, $B=[-5.0,5.0]$, $\theta=[0.001,10]$
    \item \textbf{Models 4-7}: $w_0=[-2.0, 0.0]$, $b=[-5.0, 5.0]$
    \item \textbf{Models 8-10}: $w_1=[-5.0, 5.0]$, $w_2=[-10.0,10.0]$
    \item \textbf{Models 11-12}: $w_1=[-5.0, 5.0]$, $w_2=[-3.0,3.0]$
\end{itemize}

\begin{figure*}
    \makebox[11cm][c]{
     \includegraphics[trim = 0mm  0mm 0mm 0mm, clip, width=6cm, height=5.5cm]{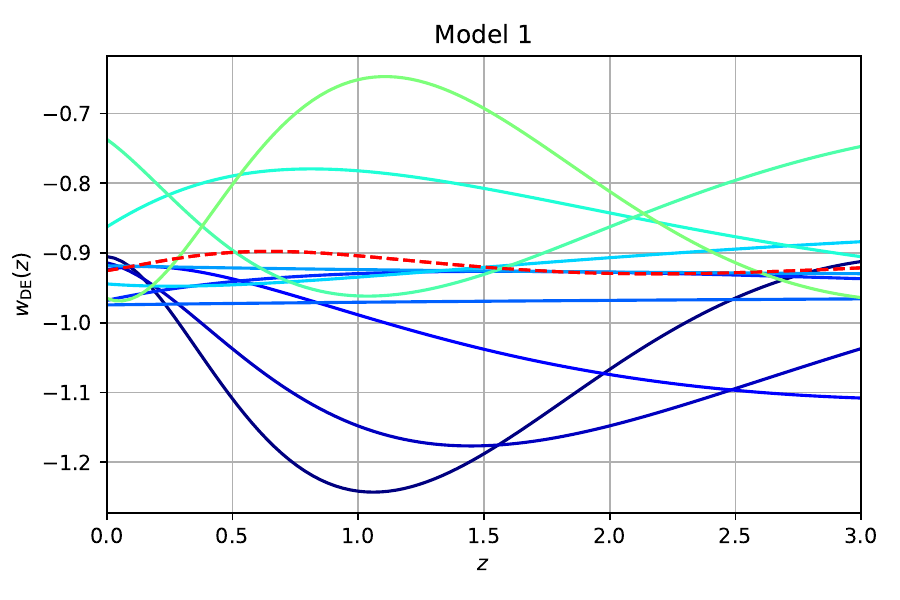}
     \includegraphics[trim = 0mm  0mm 0mm 0mm, clip, width=6cm, height=5.5cm]{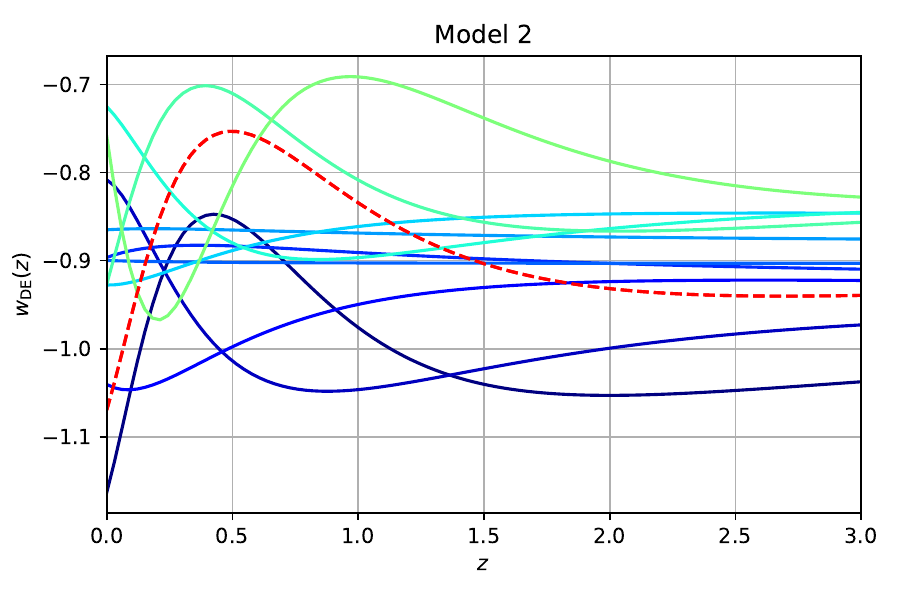}
     \includegraphics[trim = 0mm  0mm 0mm 0mm, clip, width=6cm, height=5.5cm]{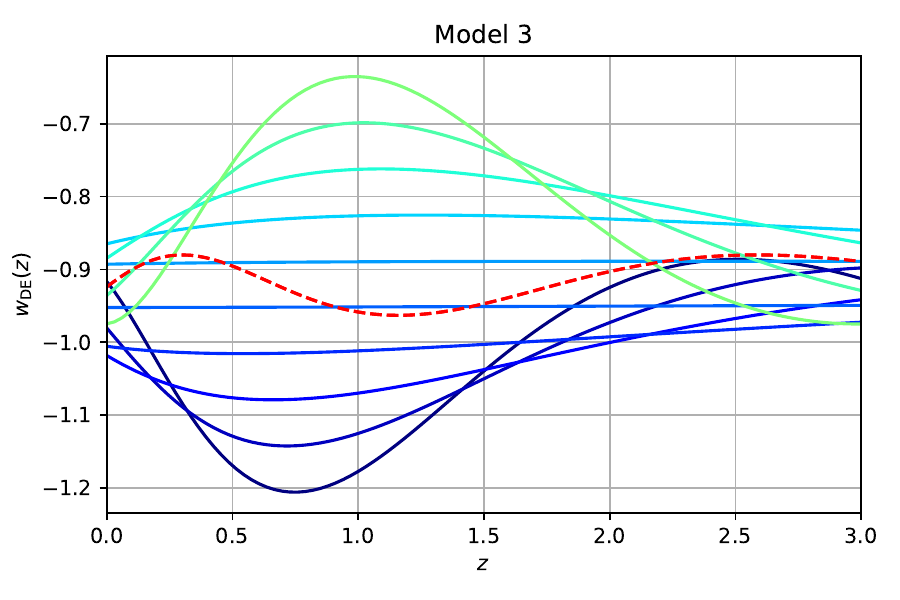}
     }
     \makebox[11cm][c]{ 
     \includegraphics[trim = 0mm  0mm 0mm 0mm, clip, width=6cm, height=5.5cm]{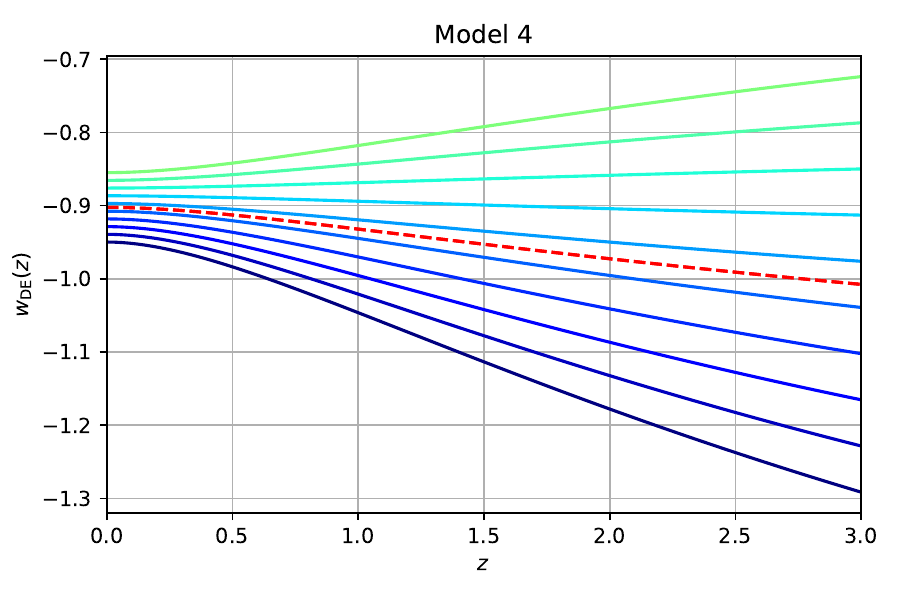}
     \includegraphics[trim = 0mm  0mm 0mm 0mm, clip, width=6cm, height=5.5cm]{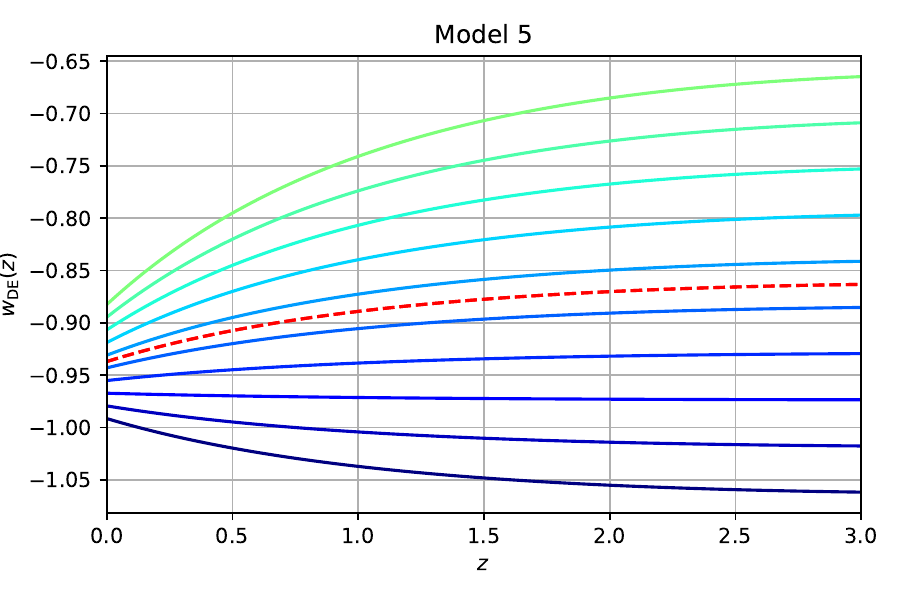}
     \includegraphics[trim = 0mm  0mm 0mm 0mm, clip, width=6cm, height=5.5cm]{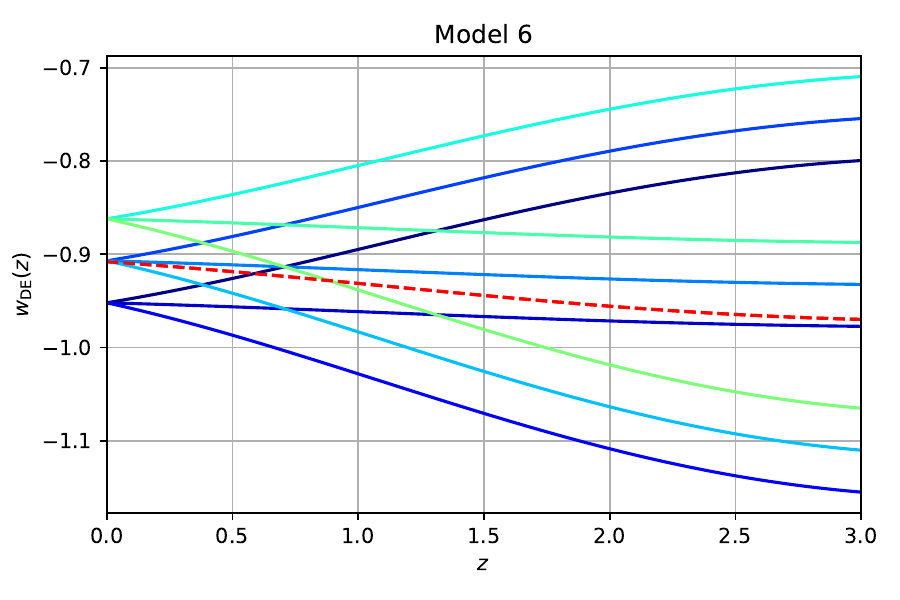}
     }
     \makebox[11cm][c]{   
     \includegraphics[trim = 0mm  0mm 0mm 0mm, clip, width=6cm, height=5.5cm]{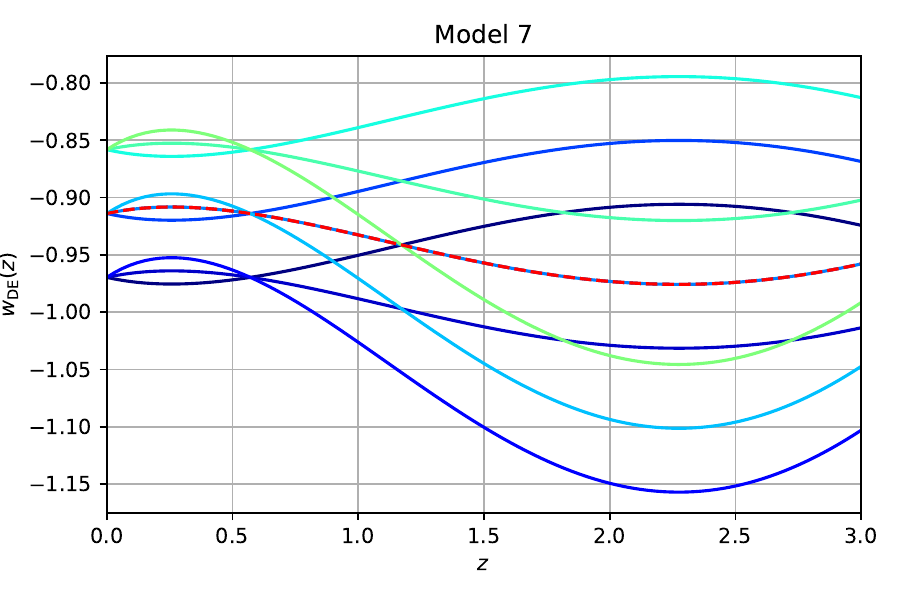}
     \includegraphics[trim = 0mm  0mm 0mm 0mm, clip, width=6cm, height=5.5cm]{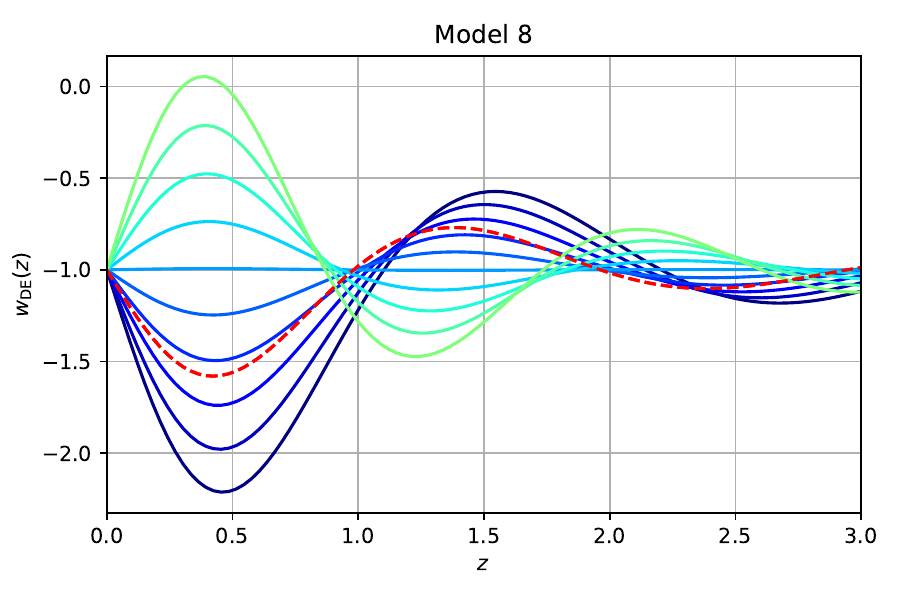}
     \includegraphics[trim = 0mm  0mm 0mm 0mm, clip, width=6cm, height=5.5cm]{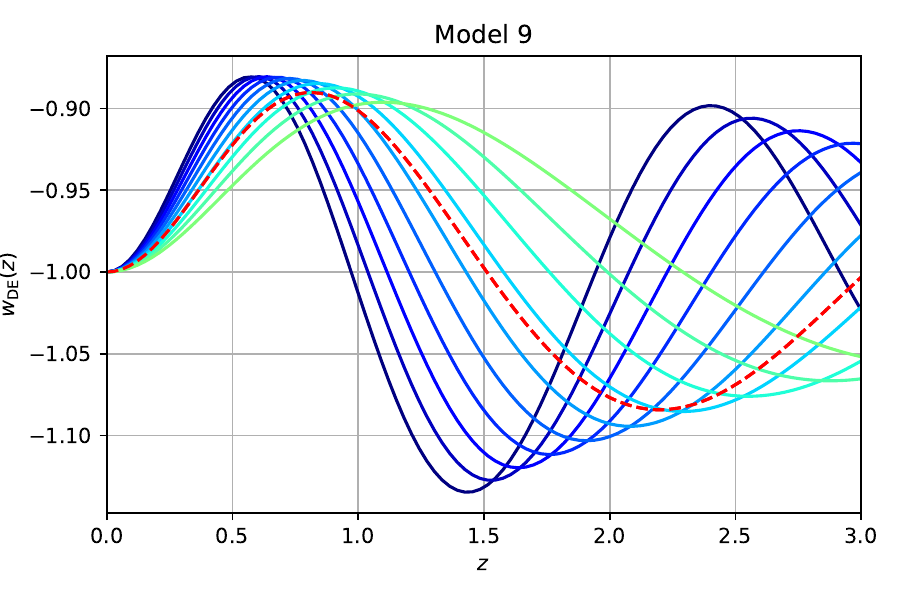}
     }
     \makebox[11cm][c]{   
     \includegraphics[trim = 0mm  0mm 0mm 0mm, clip, width=6cm, height=5.5cm]{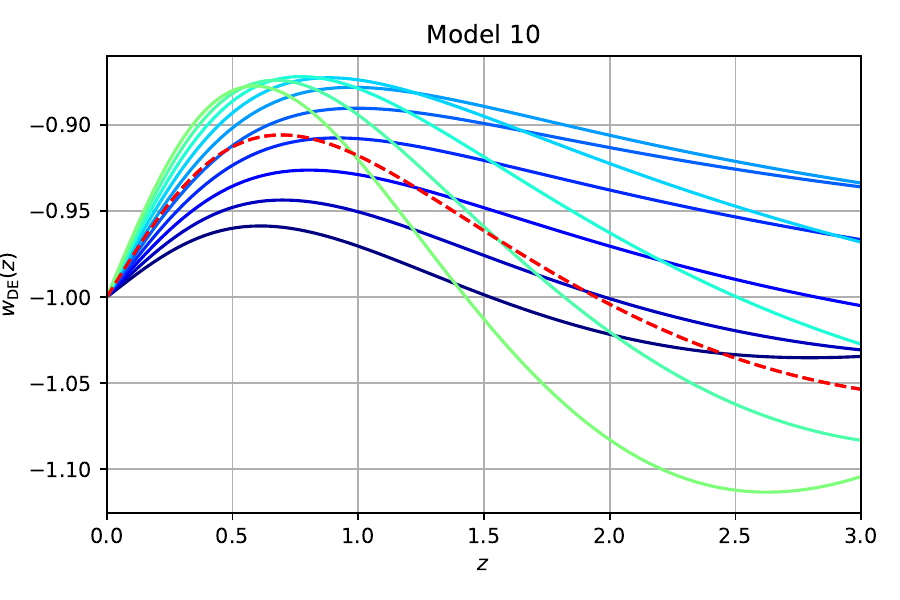}
     \includegraphics[trim = 0mm  0mm 0mm 0mm, clip, width=6cm, height=5.5cm]{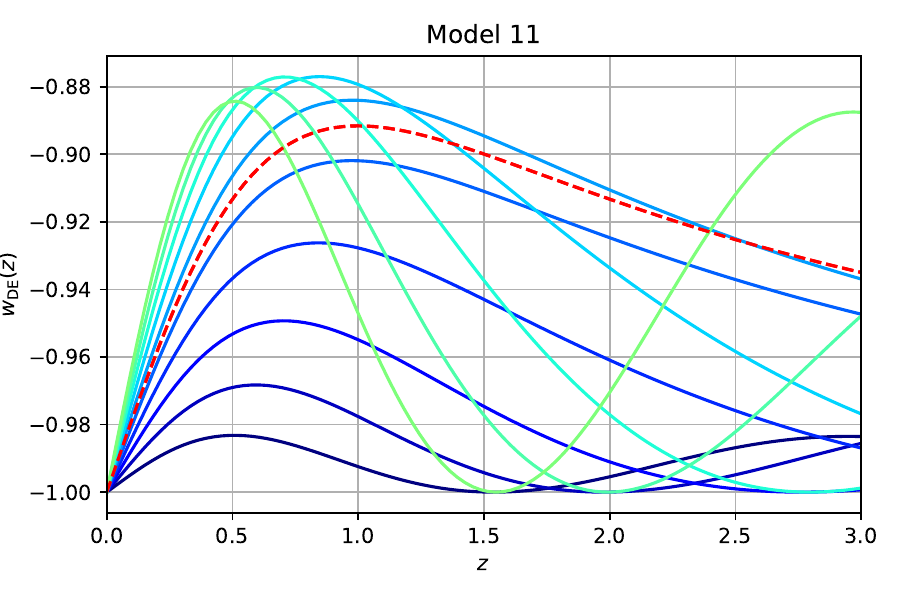}
     \includegraphics[trim = 0mm  0mm 0mm 0mm, clip, width=6cm, height=5.5cm]{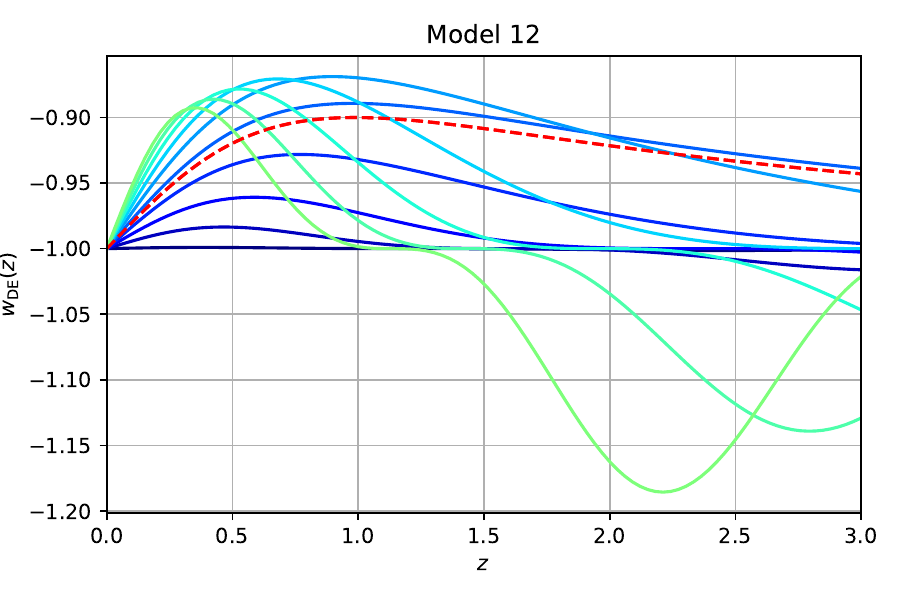}
     }
\caption{The best-fit curve (red line) together with the $1\sigma$ curves representing $w_{\rm DE}$ for all the DE parametrizations considering the combined dataset CC+SN+BAO. The different curves represent the behaviour of each parameterization when taking values distributed along the inferred $1\sigma$ range of their respective parameters.}
\label{fig:eos_1sigma}
\end{figure*}

\begin{figure*}
    \makebox[11cm][c]{
     \includegraphics[trim = 0mm  0mm 0mm 0mm, clip, width=6cm, height=5.5cm]{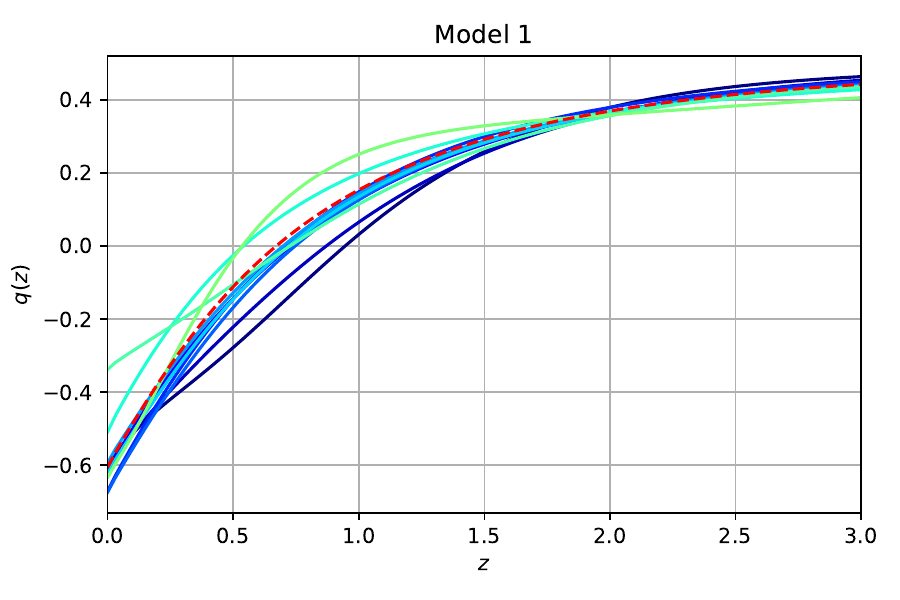}
     \includegraphics[trim = 0mm  0mm 0mm 0mm, clip, width=6cm, height=5.5cm]{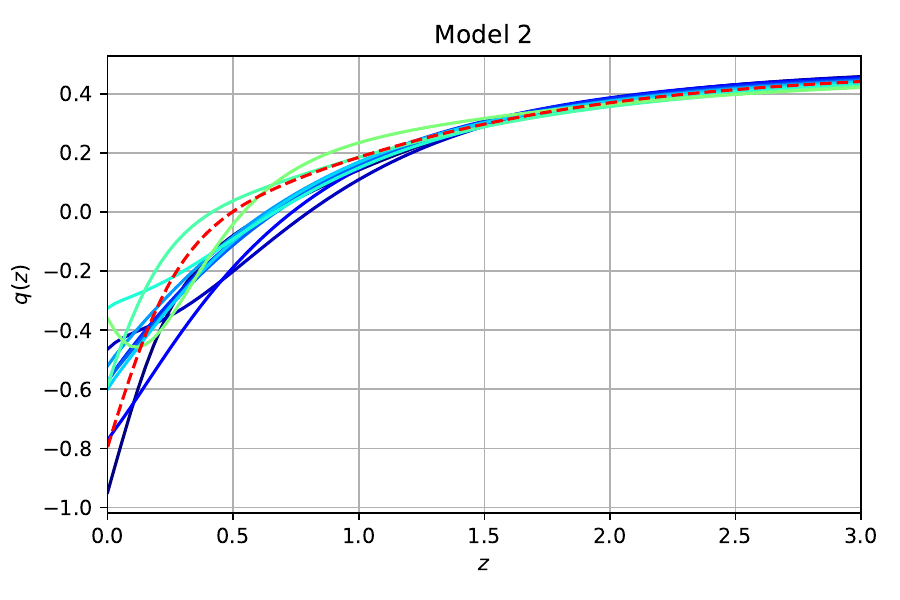}
     \includegraphics[trim = 0mm  0mm 0mm 0mm, clip, width=6cm, height=5.5cm]{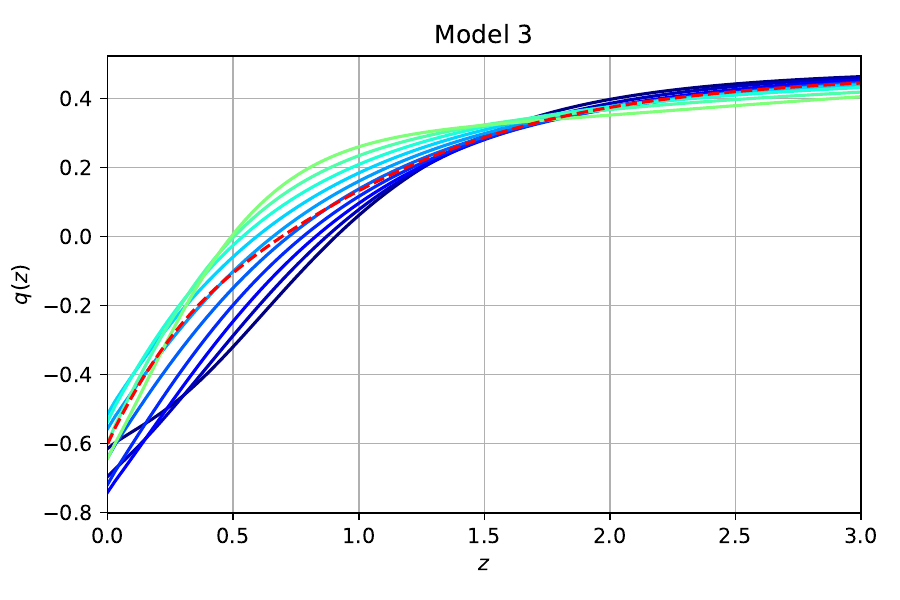}
     }
     \makebox[11cm][c]{ 
     \includegraphics[trim = 0mm  0mm 0mm 0mm, clip, width=6cm, height=5.5cm]{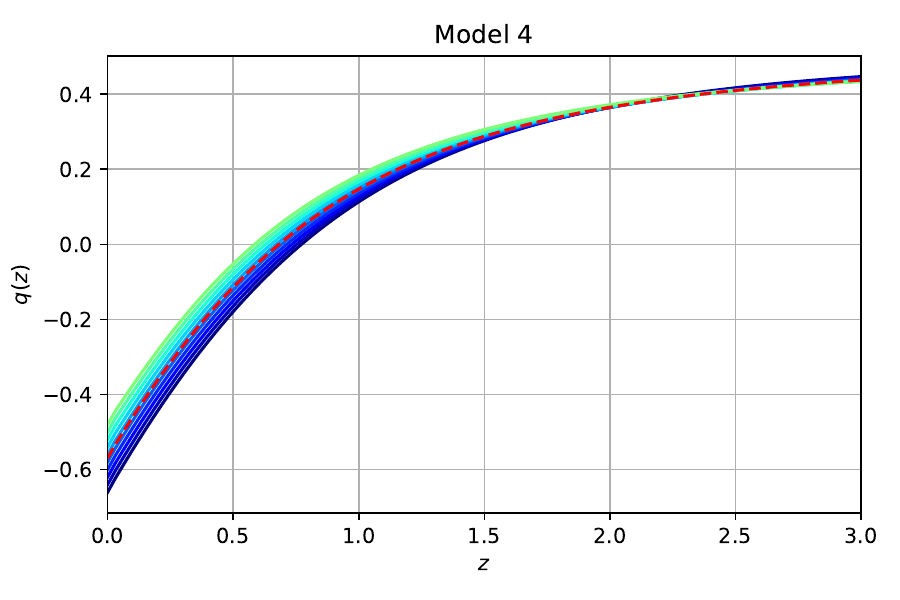}
     \includegraphics[trim = 0mm  0mm 0mm 0mm, clip, width=6cm, height=5.5cm]{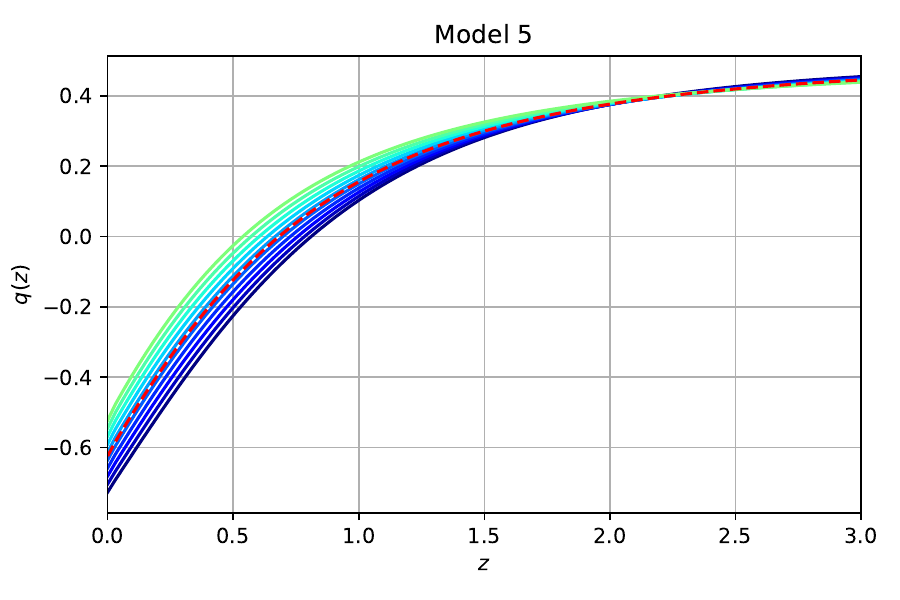}
     \includegraphics[trim = 0mm  0mm 0mm 0mm, clip, width=6cm, height=5.5cm]{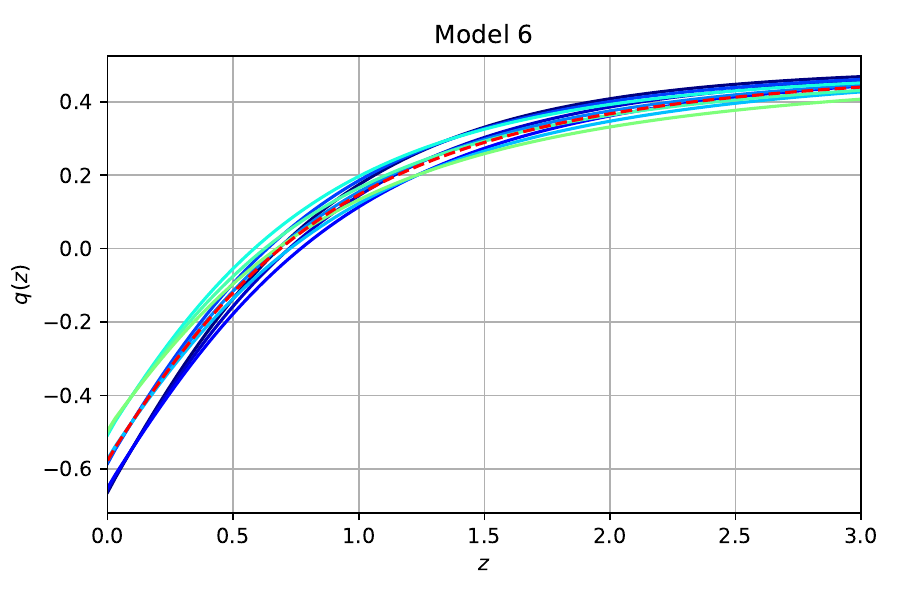}
     }
     \makebox[11cm][c]{   
     \includegraphics[trim = 0mm  0mm 0mm 0mm, clip, width=6cm, height=5.5cm]{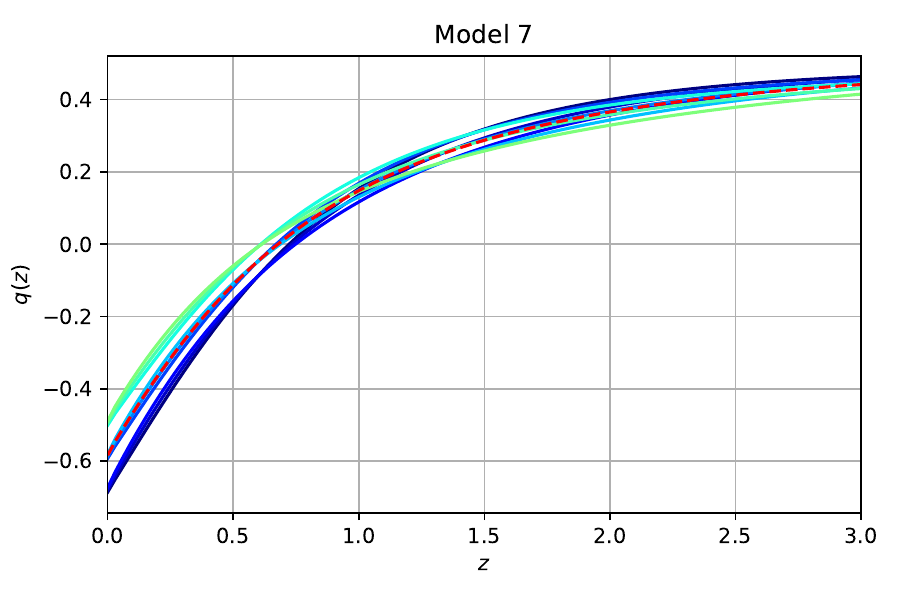}
     \includegraphics[trim = 0mm  0mm 0mm 0mm, clip, width=6cm, height=5.5cm]{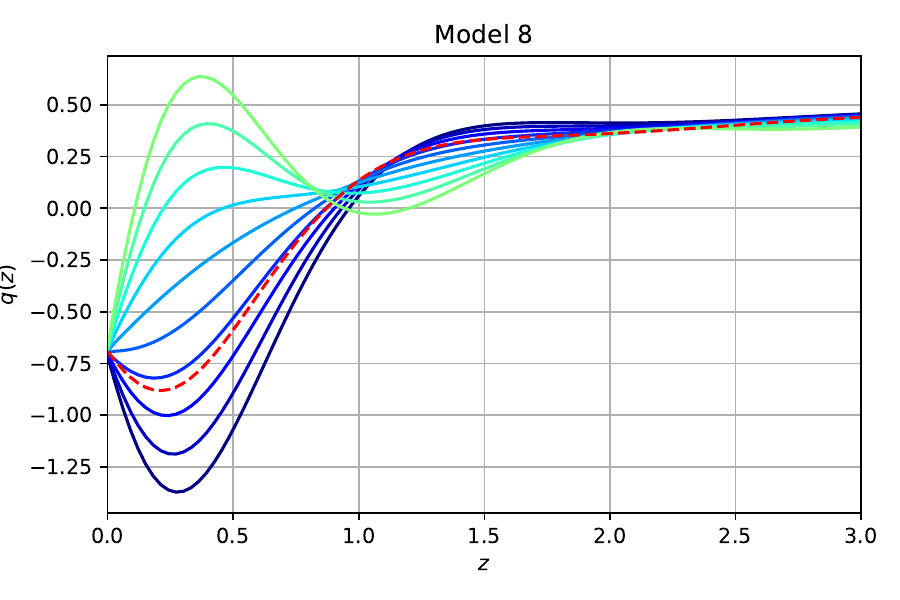}
     \includegraphics[trim = 0mm  0mm 0mm 0mm, clip, width=6cm, height=5.5cm]{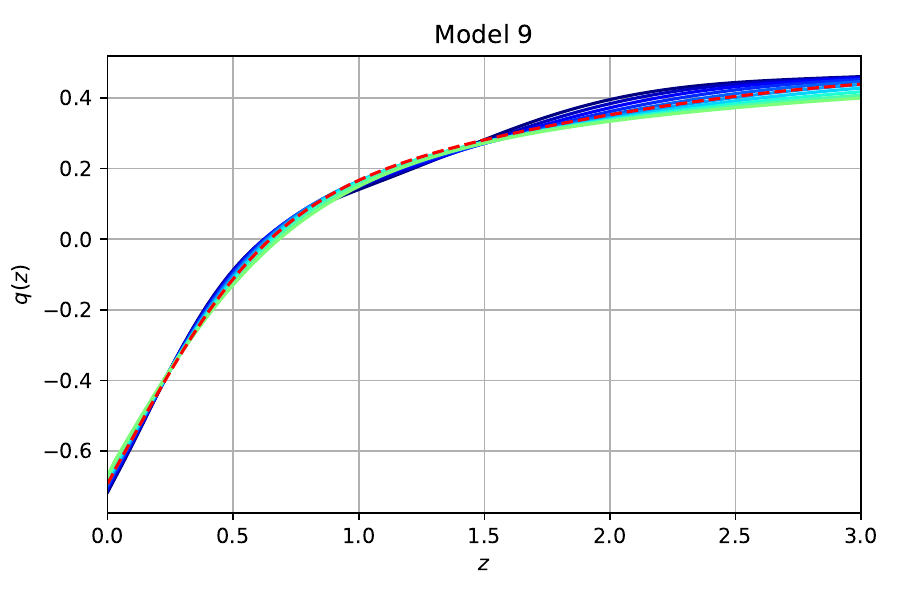}
     }
     \makebox[11cm][c]{   
     \includegraphics[trim = 0mm  0mm 0mm 0mm, clip, width=6cm, height=5.5cm]{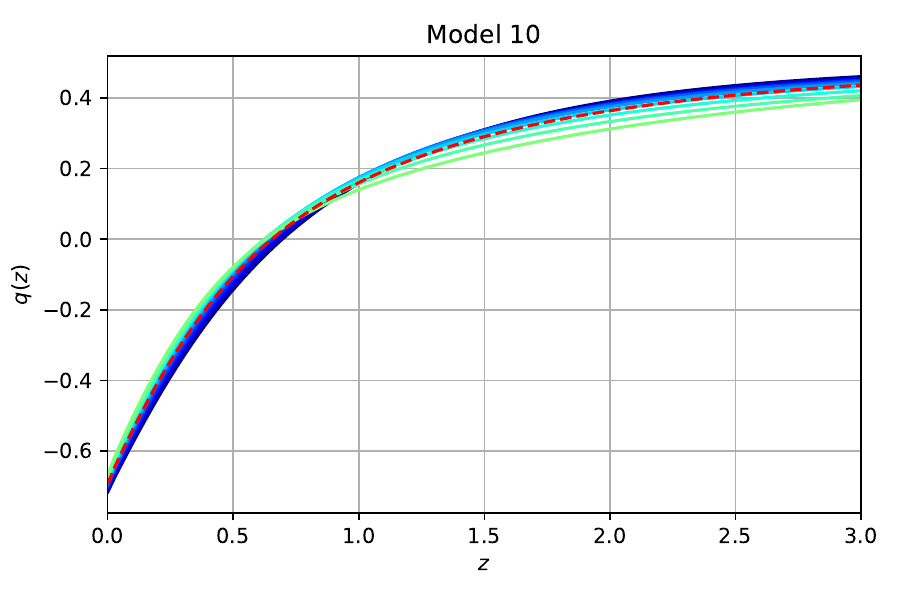}
     \includegraphics[trim = 0mm  0mm 0mm 0mm, clip, width=6cm, height=5.5cm]{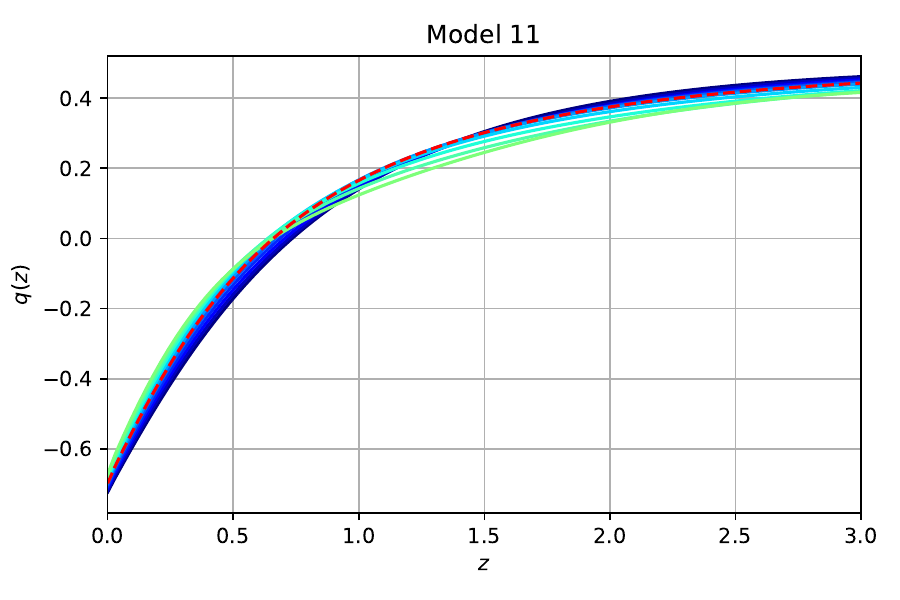}
     \includegraphics[trim = 0mm  0mm 0mm 0mm, clip, width=6cm, height=5.5cm]{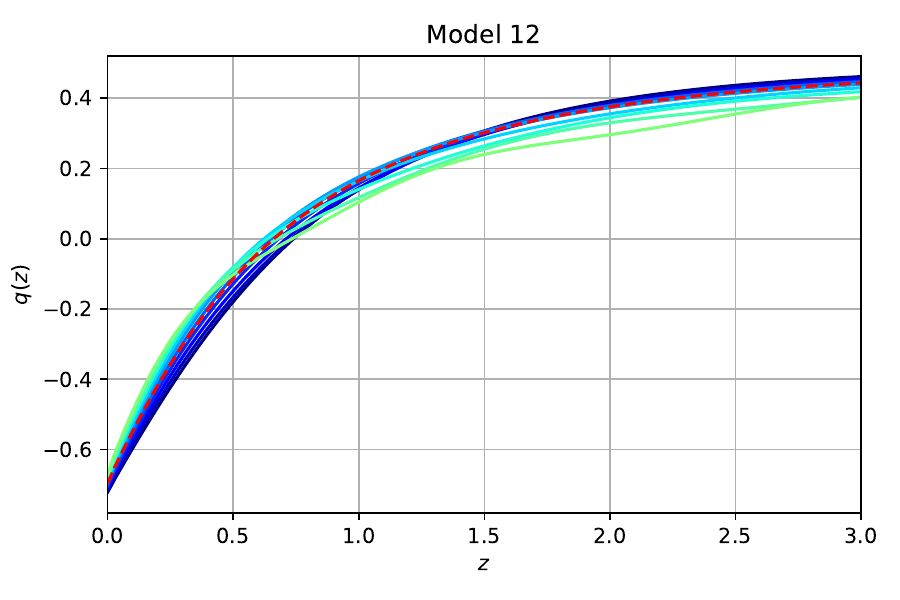}
     }
\caption{The statistical behavior of the deceleration parameter for all the  DE parametrizations has been displayed considering the combined dataset CC+SN+BAO. In each graph the red curve represents the best-fit curve filled with the $1\sigma$ region. } \label{fig:qz_1sigma}
\end{figure*}

\section{Results and their implications}
\label{sec-results}

In this section we summarize the results obtained from all the parameterizations after conducting parameter inference on them.  
We begin with the reconstructed EoS of DE using bins and CC+SN+BAO, as shown in Fig.~\ref{fig:d2comp}. From Fig.~\ref{fig:d2comp}, it is evident that the reconstructed $w_{\rm DE}$ exhibits oscillatory behavior around $w_{\rm DE} = -1$ during late times. This oscillation implies a transition between the quintessence ($w_{\rm DE} > -1$) and phantom ($w_{\rm DE} < -1$) regimes. This finding aligns with recent studies suggesting an oscillating nature of $w_{\rm DE}$ in the late-time evolution of the universe~\cite{Zhao:2017cud,Zhang:2019jsu}. We can also observe a preference for this behavior reflected in the $-2\Delta\ln \mathcal{L_{\rm max}}$ value of $-15.41$ (as shown in Table \ref{tabla_evidencias}). However, this preference can also be attributed to the high number of parameters (20) used for this reconstruction, which enables a better fit to the data. Nonetheless, these additional degrees of freedom come with a significant caveat: a penalty in the value of the Bayes' Factor, which is $4.92$ in this case, indicating it as the least favored among the options.
Now, having this reconstructed version of $w_{\rm DE}$, one can further investigate the viability of the existing oscillating DE parametrizations. However, there's an opportunity to enhance these models by utilizing fewer additional parameters to avoid substantial penalization. 

Let us now discuss the parameterizations displayed in Table~\ref{tab:models}. Their performance in fitting the data and Bayesian Evidence is reported in Table~\ref{tabla_evidencias}. The best-fit values of all the free parameters of the oscillating DE models are reported in Table~\ref{tabla:best-fit_model-params}. 
And finally, Figs. \ref{fig:eos_1sigma}
and \ref{fig:qz_1sigma} highlight their qualitative features in terms of their EoS and deceleration parameter $q(z)$. In Fig.~\ref{fig:eos_1sigma}, the evolution of all the EoS parameters is depicted. It is noticeable that, except for Model 11, all the models allow for both quintessence and phantom behavior within $1\sigma$.  
In Fig.~\ref{fig:qz_1sigma}, we present the evolution of the deceleration parameter, revealing a clear transition from the past decelerating phase to the current accelerating phase for all parametrizations. In addition, for Model 8, we observe an indication (even if not statistically significant) for the slowing down of the cosmic acceleration and based on the turning nature of the curve, maybe in future our universe could enter into a decelerating phase. 

\begin{figure}
    \centering
    \includegraphics[width=0.52\textwidth]{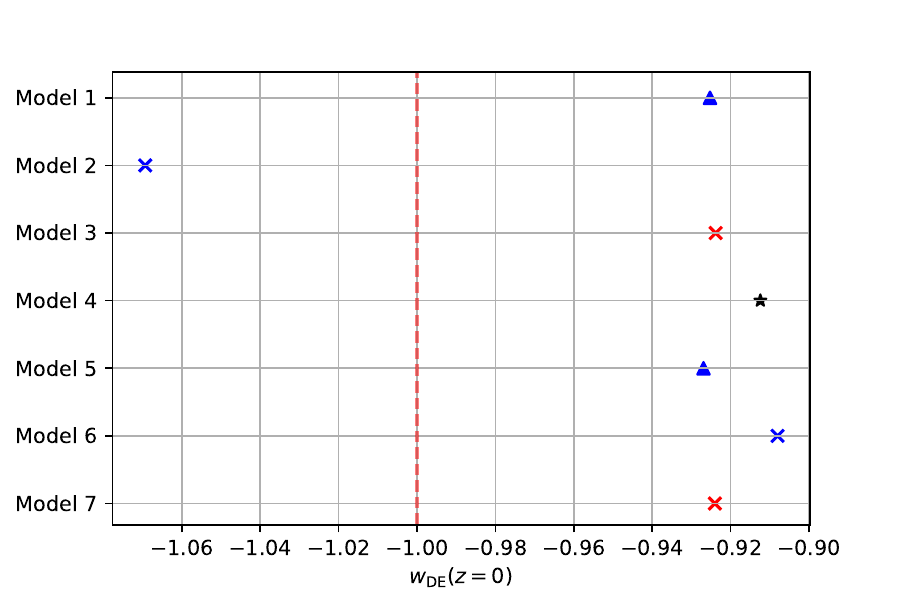}
    \caption{The best-fit value of each EoS parameterization today ($w_{\rm DE}(z=0)$). Models 8-12 cannot have an initial value different from $-1$ because of the way they are defined. The dotted vertical line corresponds to $w (z =0) = -1$.   }
    \label{fig:eos_value_today}
\end{figure}

Next in Fig.~\ref{fig:eos_value_today} we show the best-fit values of $w_{\rm DE}$ at present time (i.e. $w_{\rm DE} (z =0)$) obtained in all the EoS parameters for the combined dataset CC+SN+BAO. Note that for Models 8-12, because of the way they are defined, they attain $w_{\rm DE} (z =0) = -1$. For the remaining models, the current nature ($z=0$) of the oscillating DE EoS is quintessential in most of the cases (phantom nature only for Model 2), which is reflected from Fig.~\ref{fig:eos_value_today}. Introducing an extra parameter for models 8-12 to attain a different value at $z=0$ could potentially exhibit a similar tendency. However, we have opted to retain them as they are to maintain the number of degrees of freedom at a minimum. Nonetheless, we believe that exploring extensions to these parameterizations to address this behavior might be worthwhile in future work.

\begin{figure}
    \centering
    \includegraphics[trim = 13mm  0mm 20mm 0mm, clip, width=8.7cm, height=6.5cm]{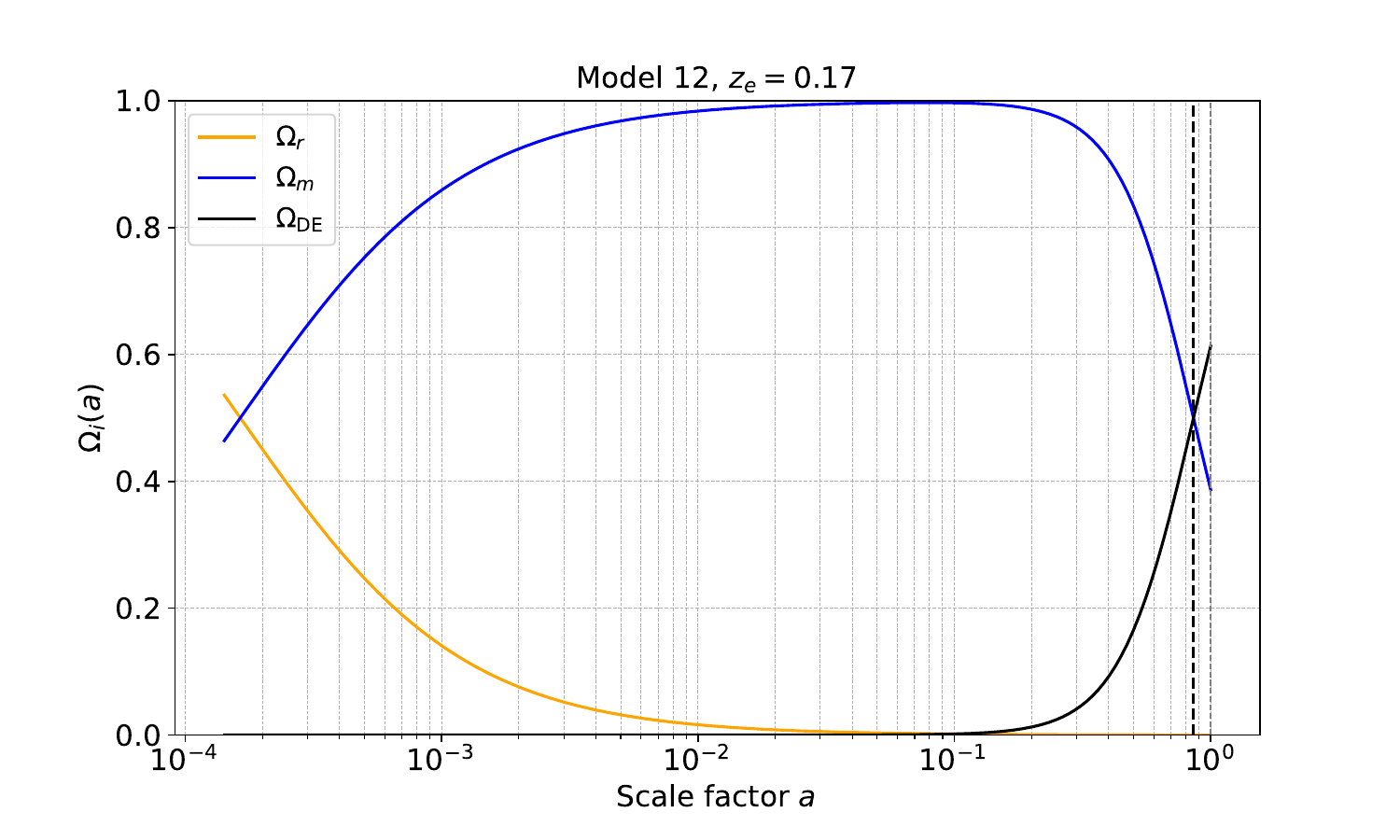}
    \caption{Evolution of the density parameters for Model 12 using its best-fit values. Although only the plot for Model 12 is shown, all other models exhibit the same general behavior. The only notable difference is the redshift value at which the transitions between domination eras occur. }
    \label{fig:density_params_12}
\end{figure}

At last in Fig.~\ref{fig:density_params_12} we present the evolution of the density parameters $\Omega_i$ for the energy components of the universe. We can see that, despite having a dynamical DE (in this case being the oscillating EoS for model 12), there are domination eras for radiation and matter\footnote{ We note that within the present oscillating DE models displayed in Table~\ref{tab:models}, our universe experiences a radiation and matter dominated eras prior to the present DE dominated era. Although we have shown these features for model 12 in Fig.~\ref{fig:density_params_12}, but one can take any of the models in Table~\ref{tab:models}. }. In the last column of Table~\ref{tabla_evidencias} we report every redshift of equivalence between matter and DE for the best-fit of every model.
In what follows we focus on each model and their responses to the observational data. 

\begin{table}[t!]
\caption{The table summarizes the mean values and the standard deviations (in parenthesis) for the parameters $h$ and $\Omega_{{\rm m}0}$ for the combined dataset CC+SN+BAO. The last two columns correspond to the Bayes Factor, which, if positive, indicates a preference for $\Lambda$CDM, and the 
$-2\Delta\ln \mathcal{L_{\rm max}} \equiv -2\ln( \mathcal{L_{\rm max}}_{,\Lambda \text{CDM}} / \mathcal{L_{\rm max,}}_i)$ evaluated for each oscillating DE parametrization which are used to compare the fit with respect to the standard $\Lambda$CDM model. Here $z_e$ denotes the redshift of equivalence between matter and DE (i.e. the redshift at which $\rho_{\rm m} =\rho_{\rm DE}$) and it has been evaluated for each model taking the best-fit values of the model parameters.
}
\footnotesize
\scalebox{0.89}{%
\begin{tabular}{cccccc} 
\cline{1-6}\noalign{\smallskip}
\vspace{0.15cm}
Model & $h$ &  $\Omega_{{\rm m}0}$  &  $\ln B_{\Lambda \text{CDM},i}$  &  $-2\Delta\ln \mathcal{L_{\rm max}}$ & $z_e$\\
\hline
 \vspace{0.15cm}
$\Lambda$CDM &   0.696 (0.017) &  0.310 (0.012)  & $-$  &  $-$ & 0.244 \\
\vspace{0.15cm}
20 bins &   0.688 (0.019) &  0.298 (0.014) & 4.92 (0.21)  &  $-15.41$ & 0.202 \\
\vspace{0.15cm}
$1$ &  0.677 (0.020) & 0.292 (0.015)  & 3.15 (0.21)  & $-4.11$ & 0.158 \\
\vspace{0.15cm}
$2$ &  0.670 (0.020) & 0.294 (0.015)  & 3.24 (0.20)  & $-4.14$ & 0.151 \\
\vspace{0.15cm}
$3$ &  0.672 (0.020) & 0.295 (0.015)  & 3.17 (0.22)  & $-4.19$ & 0.148 \\
\vspace{0.15cm}
$4$ &  0.675 (0.023) & 0.298 (0.019)  & 2.47 (0.20)  & $-3.11$ & 0.359 \\
\vspace{0.15cm}
$5$ &  0.678 (0.024) & 0.293 (0.021)  & 3.51 (0.21)  & $-3.12$ & 0.201 \\
\vspace{0.15cm}
$6$ &  0.682 (0.023) & 0.297 (0.019)  & 2.49 (0.20)  & $-3.17$ & 0.171 \\
\vspace{0.15cm}
$7$ &  0.687 (0.023) & 0.299 (0.019)  & 2.01 (0.21)  & $-3.94$ & 0.142 \\
\vspace{0.15cm}
$8$ &  0.683 (0.020) & 0.298 (0.014)  & 1.02 (0.20)  & $-4.94$ & 0.203 \\
\vspace{0.15cm}
$9$ &  0.682 (0.019) & 0.302 (0.014)  & 0.82 (0.21)  & $-5.21$ & 0.178 \\
\vspace{0.15cm}
$10$ &  0.675 (0.022) & 0.304 (0.015)  & 1.08 (0.20)  & $-5.02$ & 0.182 \\
\vspace{0.15cm}
$11$ &  0.678 (0.022) & 0.299 (0.015)  & 1.11 (0.20)  & $-4.76$ & 0.201 \\
\vspace{0.15cm}
$12$ &  0.681 (0.022) & 0.302 (0.015)  & 0.79 (0.22)  & $-4.49$ & 0.171 \\
\hline
\hline
\end{tabular}}
\label{tabla_evidencias}
\end{table}

\textbf{Models 1, 2, 3:} Models 1, 2 and 3 are some of the earliest examples of parameterizations where the EoS was allowed to oscillate, being particularly inspired by a quintom DE (a DE whose EoS can cross the phantom divide line $w_{\rm DE}=-1$). Given that these three models have the largest number of free parameters, i.e. 4, it is unexpected that they are not the best fit to the data (when compared to the other models), which is reflected in their $-2\Delta\ln \mathcal{L_{\rm max}}$ values as $-4.11$, $-4.14$ and $-4.19$, respectively. We attribute this to the fact that, according to~\cref{fig:d2comp}, the data prefers oscillations at late times, which these models have problems fulfilling due to the way they are defined (their oscillatory behavior becomes more prominent at high redshifts). Unsurprisingly, though, their Bayes' factors are higher than most of the other models with fewer free parameters, presenting moderate evidence in favor of $\Lambda$CDM.

\textbf{Models 4, 5, 6:} This triad of models shows some improvement in fitting the data, although it is the smallest among the models studied. This limitation arises from a significant drawback: they lack a parameter to adjust the frequency of oscillations. Consequently, during a parameter inference procedure, their oscillations remain static. While these models indeed exhibit oscillatory behavior, it occurs primarily at early times (similar to models 1, 2 and 3) rather than at late times (low redshifts). Introducing a new parameter to modify the oscillation frequency would be necessary to address this issue. However, it is worth noting that such an addition would likely result in a worse Bayesian Evidence, and they already have moderate evidence against them when compared to the standard model.

\textbf{Model 7:} Similar to Models 4, 5, and 6, Model 7 shares a disadvantage: lacking a parameter to modulate the oscillation frequency. This is apparent when looking at its $-2\Delta\ln \mathcal{L_{\rm max}}$ value of $-3.94$, making it the fourth worst among the group. However, it does exhibit a crossing of the phantom divide line  at low redshift. The Bayes' factor aligns with expectations, considering its number of parameters and the similarity in their priors with the preceding three models. 

\textbf{Models 8, 9, 10:} Models 8 and 9 exhibit the most notable oscillatory behavior, evident at the $1\sigma$ level, as illustrated in Fig~\ref{fig:eos_1sigma}. Model 8 demonstrates the ability to cross the phantom divide line up to 3 times, while model 9 achieves up to 2 crossings. Notably, Model 9 boasts the best fit among all the models studied in this work. Their Bayes' factors provide moderate evidence against themselves, in line with expectations, given the presence of 2 extra parameters and relatively broad priors compared to other models. Model 10 shares some visual similarities with models 7 and 9, with the latter also presenting a similar Bayes' factor albeit a worse fit to the data (but not by much).

\textbf{Models 11, 12:} These last two entries show the most favorable Bayes' factors in the comparison, although they still fall short compared to the standard model's one, exhibiting only weak evidence against them according to the revised Jeffreys' scale. This is likely due to the narrower prior on the parameter $w_2$ compared to Models 4-10. Their fit to the data is moderate, and this might be attributed to certain characteristics of these parameterizations: Model 11, by design, cannot cross the phantom divide line, a behavior favored by every other reconstructed EoS in this work, putting it at a disadvantage; Model 12, featuring a cubic cosine function, introduces some flatness in the oscillations at certain intervals, attempting to identify this characteristic in Fig~\ref{fig:d2comp} (which is model-independent) will prove to be a futile task, making this particular behavior less preferred when using these datasets to perform parameter inference.

\begin{table}[t!]
\caption{The table summarizes the mean values and the standard deviations (in parenthesis when symmetric) for the model-specific parameters. 
}
\footnotesize
\scalebox{0.89}{%
\begin{tabular}{ccccc} 
\cline{1-5}\noalign{\smallskip}
\vspace{0.15cm}
Model &   &   &   &  \\
\hline
\vspace{0.15cm}
     & $w_0$ &  $w_1$  &   $A$  &  $a_c$ \\
\vspace{0.15cm}
$1$ &  $-0.98$ (0.09) & 0.02 (0.21) &  0.98 (4.23)  & 4.99 (2.85) \\
\vspace{0.15cm}
    & $w_1$ &  $w_2$  &   $w_3$  &  $a_c$ \\
$2$ &  $-0.91$ (0.11) & 0.002 (0.207) &  $-0.304$  (7.134)  & 5.18 (2.82) \\
\vspace{0.15cm}
    & $w_c$ &  $A$  &   $B$  &  $\theta$ \\
$3$ &  $-0.97^{+0.19}_{-0.08}$ & $-0.041^{+0.18}_{-0.15}$ &  $6.21^{+0.28}_{-4.09}$  & $6.31~(2.81)$ \\
\hline 
\vspace{0.15cm}
    & $w_0$ &  $b$  &  $-$  & $-$ \\
    \hline 
\vspace{0.15cm}

$4$ &  $-0.911~(0.043)$ & $-0.111~(0.279)$ & $-$  & $-$   \\
\vspace{0.15cm}
$5$ &  $-0.928~(0.057)$ & $0.072~(0.155)$ &  $-$ &  $-$ \\
\vspace{0.15cm}
$6$ &  $-0.911~(0.049)$ & $0.058~(0.189)$ & $-$  &  $-$ \\
\vspace{0.15cm}
$7$ &  $-0.921~(0.054)$ & $0.088~(0.191)$ & $-$  &  $-$ \\
\hline 
     & $w_1$ &  $w_2$  &  $-$ &  $-$ \\
     \hline 
\vspace{0.15cm}
$8$ &  $-0.72^{+1.93}_{-0.81}$ & $3.29^{+0.42}_{-0.31}$ &  $-$ & $-$  \\
\vspace{0.15cm}
$9$ &  $-0.23^{+0.02}_{-0.04}$ & $-2.08^{+0.51}_{-1.03}$ &  $-$ & $-$  \\
\vspace{0.15cm}
$10$ &  $0.24^{+0.11}_{-0.12}$ & $-0.83^{+1.21}_{-0.32}$ & $-$  & $-$  \\
\vspace{0.15cm}
$11$ &  $0.209~(0.169)$ & $-0.005~(1.101)$ &  $-$ & $-$ \\
\vspace{0.15cm}
$12$ &  $0.189~(0.195)$ & $0.067~(1.231)$ & $-$  &  $-$ \\
\hline
\hline
\end{tabular}}
\label{tabla:best-fit_model-params}
\end{table}

\section{Summary and Conclusions}
\label{sec-summary}

With the abundances of astronomical data and the availability of sophisticated numerical packages, cosmology has become increasingly exciting in the twenty-first century. Understanding the dynamics of the universe has emerged as a pivotal challenge for modern cosmology over the last several years. While the standard $\Lambda$CDM cosmology has been successful in explaining numerous astronomical surveys, this century has highlighted that $\Lambda$CDM may not be the ultimate theory of the universe. Discrepancies between early and late cosmological probes, with the $H_0$ tension being a major challenge, suggest the need for alternative explanations.

Tracing the evolutionary history of the universe, commonly two approaches are considered. The simplest and most commonly used approach, as documented in existing literature, is to propose a cosmological model and assess how well available astronomical data aligns with this model. This method enables the introduction of a sequence of cosmological models with the aim of identifying the most optimal one based on the observational datasets.
Alternatively, using the available astronomical probes, one can reconstruct the expansion history of the universe through cosmological variables. For instance, by employing a suitable numerical algorithm, one can reconstruct the Hubble parameter and its derivatives. Subsequently, a cosmological parameter expressed as a function of these variables can be reconstructed. Unlike the previous approach, this method is model-independent, as it does not require the assumption of an arbitrary parametrization.

In seeking to understand how the model-independent approach influences the evolution of DE, we have noticed that $w_{\rm DE}$ may exhibit an oscillating nature between quintessence and phantom states (see Fig.~\ref{fig:d2comp}) and this result is in agreement with earlier reports by other authors~\cite{Zhao:2017cud,Zhang:2019jsu}. 
This served as an inspiration to revisit oscillating DE parametrizations previously introduced in the literature by numerous investigators~\cite{Rubano:2003er,Linder:2005dw,Feng:2004ff,Nojiri:2006ww,Kurek:2007bu,Jain:2007fa,Saez-Gomez:2008mkj,Kurek:2008qt,Pace:2011kb,Pan:2017zoh,Panotopoulos:2018sso,Rezaei:2019roe,Guo:2022dno}. 
In this article, we have considered 12 oscillating DE EoS parameterizations (see Table~\ref{tab:models}), among which 7 models have been previously proposed in earlier works, while 5 models are newly introduced. The aim of this article is to determine which of these 12 distinct models offer the best approximation to the reconstructed DE EoS as in  Fig.~\ref{fig:d2comp}.
Among all of these, we find that Models 8 and 9 exhibit the most remarkable oscillatory features (see Fig.~\ref{fig:eos_1sigma}). Specifically, for Model 8, $w_{\rm DE}$ crosses the phantom divide line $w_{\rm DE} = -1$ up to 3 times, while for Model 9, 2 crossings of the phantom divide line are observed. Hence, these two models can be considered to have similar (though not identical) features as depicted in Fig.~\ref{fig:d2comp}.

On the other hand, concerning the fit to the data, our analyses clearly demonstrate that every EoS considered in this work fits the data significantly better compared to the $\Lambda$CDM model, as quantified through $-2\Delta\ln \mathcal{L_{\rm max}}$ (see the \textbf{fifth} column of Table~\ref{tabla_evidencias}). However, in terms of Bayesian evidence analysis (see the fourth column of Table~\ref{tabla_evidencias}), none of the models are preferred over $\Lambda$CDM. This is because the present oscillating DE parameterizations have 2-4 extra free parameters compared to the $\Lambda$CDM model, resulting in a higher number of degrees of freedom. While having more degrees of freedom generally leads to a better fit, it also entails being penalized by the Bayes' Factor. \textbf{Nevertheless, it should be mentioned that, among all the oscillating DE models studied here, Models 9 and 12 deserve special attention given that they are the only ones which obtained a $\ln B_{\Lambda \text{CDM},i}<1$, with Model 12 having the best Bayes' factor overall. Qualitatively both models exhibit oscillating features at low redshifts, with Model 12 also manifesting some flatness at certain intervals. Those distinguishing features should be paid more attention to in light of these findings.}

In summary, this study aims to identify a DE EoS capable of replicating the behavior of the model-independent reconstruction of $w_{\rm DE}$ (Fig.~\ref{fig:d2comp}) up to $z =3$, thereby incorporating low redshift cosmological probes into the analysis. It is important to note that i) high redshift cosmological probes have not been considered in the reconstruction, and ii) upcoming astronomical surveys are expected to introduce more potential cosmological probes in the coming years. As a result, the final conclusion in this regard may hold surprises for future investigations.

\section*{Acknowledgments}
The authors thank the reviewer for some insightful comments which improved the quality of the article.
LAE acknowledges support from the Consejo Nacional de Humanidades, Ciencias y Tecnolog\'{i}as (CONAHCyT, National Council of Humanities Science and Technology of Mexico) and from the Programa de Apoyo a Proyectos de Investigación e Innovación Tecnológica (PAPIIT) from UNAM IN117723.  SP acknowledges the financial support from the Department of Science and Technology (DST), Govt. of India under the Scheme   ``Fund for Improvement of S\&T Infrastructure (FIST)'' (File No. SR/FST/MS-I/2019/41). EDV acknowledges support from the Royal Society through a Royal Society Dorothy Hodgkin Research Fellowship. 
AP thanks UCN for the support by the Resoluci\'{o}n VRIDT No. 098/2022 and the Núcleo de Investigación Geometría Diferencial y Aplicaciones, Resolución VRIDT N°096/2022. AP thanks Nikolaos Dimakis and the Universidad de La Frontera for the hospitality provided when this work was carried out. 
JAV acknowledges the support provided by FOSEC SEP-CONACYT Investigaci\'on B\'asica A1-S-21925, FORDECYT-PRONACES-CONACYT/304001/2020 and UNAM-DGAPA-PAPIIT IA104221. WY has been  supported by the National Natural Science Foundation of China under Grants No. 12175096, and Liaoning Revitalization Talents Program under Grant no. XLYC1907098. 
We also want to thank the Unidad de C\'omputo of ICF-UNAM for their assistance in the maintenance and use of the computing equipment.
This article is based upon work from the COST Action CA21136 ``Addressing observational tensions in cosmology with systematics and fundamental physics (CosmoVerse), supported by COST (European Cooperation in Science and Technology).

\bibliography{biblio}
\end{document}